\documentclass{svjour2}                    
\smartqed  
\usepackage{graphicx}
\usepackage{subfigure}
\usepackage{ulem}
\usepackage{dcolumn}
\usepackage{bm}
\usepackage{amsfonts}
\usepackage{amsmath}
\usepackage{amssymb}

\begin{document}

\newcommand{\be}{\begin{equation}}
\newcommand{\ee}{\end{equation}}
\newcommand{\bea}{\begin{eqnarray}}
\newcommand{\eea}{\end{eqnarray}}
\newcommand{\bse}{\begin{subequations}}
\newcommand{\ese}{\end{subequations}}
\newcommand{\comment}[1]{}
\newcommand{\eps}{\varepsilon}

\title{Coarsening dynamics in a simplified DNLS model}


\author{Stefano Iubini \and Antonio Politi \and Paolo~Politi}

\institute{ Stefano Iubini 
           \at  Dipartimento di Fisica e Astronomia - CSDC, Universit\`a di Firenze
	   , via G. Sansone 1 I-50019, Sesto Fiorentino, Italy\\
	   \email{stefano.iubini@fi.isc.cnr.it}   
	   \and
	   Antonio Politi 
           \at Institute for Complex Systems and Mathematical Biology \& SUPA
	   University of Aberdeen, Aberdeen AB24 3UE, United Kingdom\\ 
	   \email{a.politi@abdn.ac.uk}
	   \and
           Paolo Politi \and Stefano Iubini \at
           Istituto dei Sistemi Complessi, Consiglio Nazionale
           delle Ricerche, via Madonna del Piano 10, I-50019 Sesto Fiorentino, Italy\\
           INFN Sezione di Firenze, 
	   via G. Sansone 1 I-50019, Sesto Fiorentino, Italy\\
           \email{Paolo.Politi@isc.cnr.it}
}

\date{Received: date / Accepted: date}

\maketitle

\begin{abstract}
We investigate the coarsening evolution occurring in a simplified stochastic model 
of the Discrete NonLinear Schr\"odinger (DNLS) equation in the so-called 
negative-temperature region. We provide an explanation of the coarsening exponent 
$n=1/3$, by invoking an analogy with a suitable exclusion process. In spite of
the equivalence with the exponent observed in other known universality classes,
this model is certainly different, in that it refers to a dynamics with two
conservation laws.
\end{abstract}

\section{Introduction}

The DNLS model~\cite{Kevrekidis} (with a positive defined quartic term) is known to be characterized 
by a so-called negative temperature region, where localized solutions (breathers)
spontaneously arise and survive for long times~\cite{Flach2008,Rasmussen2000}.
In a series of theoretical 
papers, Rumpf~\cite{Rumpf2004,Rumpf2007,Rumpf2008,Rumpf2009} has shown, with the help
of entropic arguments, that the system is expected to converge towards a state 
characterized by a single breather sitting on a homogeneous background at infinite
temperature. Nevertheless, a detailed numerical study~\cite{iubini2013} has recently
challenged such a conclusion, since the process turns out to be both extremely slow and
accompanied by the continuous birth and death of breathers.

A clarification of the asymptotic regime in the original DNLS model is a rather ambitious 
task, because of both its nonlinear character and the weak coupling between breathers and
background. In order to overcome such difficulties, a purely stochastic Microcanonical 
Monte Carlo (MMC) model was proposed in Ref.~\cite{iubini2013} with the goal 
of exploring the role of entropy not only in the identification of the asymptotic state, 
but also for the characterization of the convergence process. Analogously to the DNLS
equation, the MMC model is characterized by two conservation laws (energy and norm)
and { by a local evolution rule. More precisely, such constraints are 
representative of the original DNLS dynamics in the high-norm density limit,
where the interaction energy between neighbouring sites
is negligible.}
\comment{the (stochastic) evolution rule is local. In practice, the MMC model provides a fair
description of the DNLS in the high-norm density limit, where the interaction energy 
is negligible.}
In Ref.~\cite{iubini2013} it was discovered that the MMC dynamics is
characterized by a coarsening process during which the localized solutions progressively 
disappear, while their typical height increases. On the one hand, the purely stochastic 
character of the MMC model makes it doubtful that the resulting evolution is able to 
reproduce the key features of the DNLS dynamics. On the other hand, its study can help
to clarify the kind of constraints that are expected to emerge during the convergence
to the asymptotic state.

Coarsening is a fairly common feature of out-of-equilibrium systems, either relaxing 
towards equilibrium~\cite{Bray} or kept well far from equilibrium~\cite{PREcoarsening}.
It corresponds to a growth of the typical length scale $L$ of the system, which 
usually increases with a power law, $L(t) \approx t^n$, which defines the coarsening 
exponent $n$.  According to the value of $n$, different universality classes can be defined,
which depend, first of all, on conservation laws. 
{ Two exponents, $n=1/2$ and $n=1/3$, are specially widespread. They correspond to diffusive
processes with ($n=1/3$) and without ($n=1/2$) the conservation of the order parameter.
The simplest models displaying such coarsening behaviour are the so-called models A and B of 
dynamics~\cite{review_HH}, which represent, for example, the relaxation to equilibrium of a
deeply quenched Ising model, when the magnetization is not conserved (model A) or
is conserved (model B) by dynamics.
The latter case, because of the equivalence between Ising and lattice gas model, makes the
exponent $n=1/3$ of special relevance for condensation processes and for large classes of
nonequilibrium statistical mechanics models~\cite{zrp_review}.}
\comment{
but also on certain features of the
the potential, in the case a Lyapunov functional can be defined.
}

The MMC model is still too complicated to be able to predict the universality class 
it belongs to. In fact, it is not even possible to anticipate its qualitative dynamics:
the fact it produces coarsening is not trivial at all. Once noticed that simulations do show 
coarsening, it is natural to expect an exponent $n$ strictly smaller than ${1\over 2}$, 
because conservation laws always slow down coarsening. In fact, we will find $n={1\over 3}$, 
a very common value for systems where the order parameter is conserved. However, our model
has two conserved quantities and cannot be mapped to any known model. Therefore, this is 
not a standard result. Rather, it highlights certain properties which might be common to 
more complicated systems.

The paper is organized as follows. In section 2, we introduce the model and show some general 
properties of its evolution, when the initial condition is chosen within the negative 
temperature region. In section 3 we illustrate the coarsening process for generic initial 
conditions, with particular reference to the scaling behavior of the number of surviving 
breathers. The following two sections are devoted to a discussion of two regimes: 
the fast relaxation which drives the background towards an infinite-temperature state 
(section 4), the slow relaxation that is responsible for the coarsening process (section 5).
Such studies help to identify the reason for the increasing slowness of the coarsening,
that is then better clarified in section 6, thanks to an analogy with a suitable
exclusion process. By combining the various elements, a justification for the exponent
$n=1/3$ is given.
Finally, in section 7, we summarize the main results and mention the still open problems.

\section{The model}

\begin{figure}
\hfill
\subfigure{\includegraphics[width=5 cm]{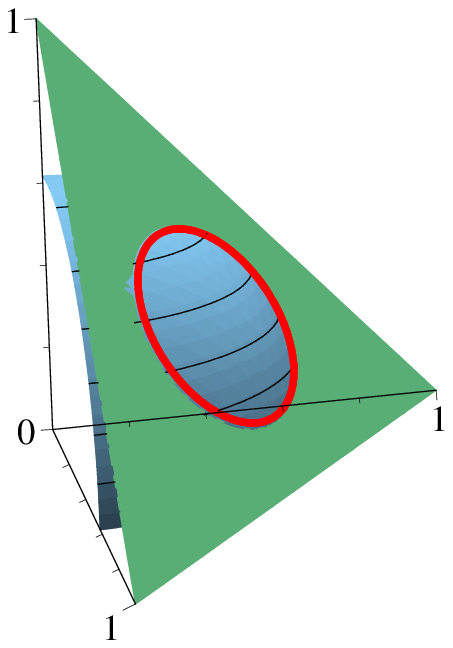}}
\hfill
\subfigure{\includegraphics[width=5 cm]{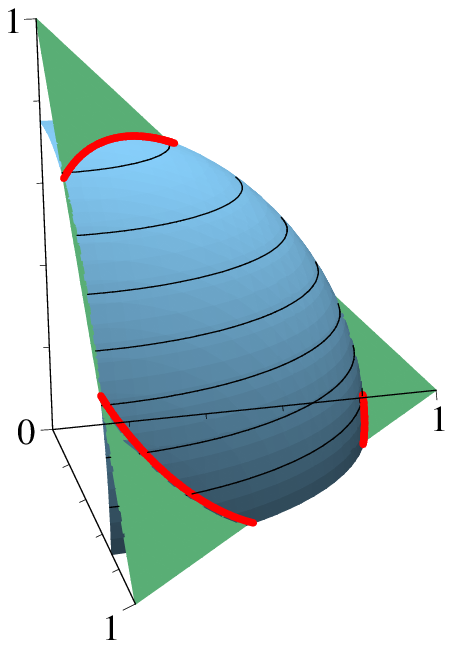}}
\hfill
\caption{
Accessible { MMC} states (red thick line) \comment{in $\mathcal{R}$} as intersection between the plane
$a_{i-1}+a_i+a_{i+1}=1$ and a sphere with square radius 0.4 (left panel) and 0.58 (right panel).
}
\label{fig_schema}
\end{figure}

In this section we define the model and discuss some general properties of its dynamical behavior.
A positive amplitude $a_i$ is defined on each site of a lattice of length $L$, where periodic boundary 
conditions are assumed. Two quantities are conserved during the evolution, namely the amplitude (mass)
\be
A = \sum_i a_i ,
\ee
and the energy 
\be
H = \sum_i a^2_i ;
\ee
which correspond to norm and energy in the DNLS equation in the large norm-density limit~\cite{iubini2013}.
The MMC model is defined as follows. Given a generic configuration at time $t$, 
a triplet $(i-1,i,i+1)$ of consecutive sites is randomly selected and updated so as to ensure that
the mass and energy are locally (and thereby globally) conserved, 
\bse
\be
a_{i-1}(t+1) + a_i(t+1) + a_{i+1}(t+1) =
a_{i-1}(t) + a_i(t) + a_{i+1}(t) 
\ee
\be
a^2_{i-1}(t+1) + a^2_i(t+1) + a^2_{i+1}(t+1) =
a^2_{i-1}(t) + a^2_i(t) + a^2_{i+1}(t) .
\ee
\label{dyn_rule}
\ese
These two laws are basically equivalent to the conservation of momentum and energy in a 
chain of oscillators (once $a_i$ is interpreted as the oscillator velocity) and one might thus
expect similar diffusion phenomena. Here, there is, however, an additional constraint: all $a_i$ 
must be positive.
Therefore, the legal configurations, that are located along the circle identified as the intersection 
between a plane and a sphere (the above two conditions), may be further confined to three separate
arcs. This happens when the maximal amplitude $\overline a$ (within the triplet) is sufficiently larger
than the other two amplitudes (see Fig.~\ref{fig_schema}). If, for simplicity, $\underline a$ is the
(equal) amplitude of the two other sites, it is easy to check that a ``three-arcs" solution 
appears when ${\overline a} > 4 {\underline a}$.  

Whenever this is the case, the definition of the model is completed by specifying that we restrict the
choice to the same arc of the initial condition. This choice is motivated by the will to
reproduce as closely as possible the original DNLS dynamics. In fact, the most important instances
of three-arcs solutions are ``breathers", i.e. isolated sites with an anomalously large amplitude.  
Such breathers, once generated, do not appear to diffuse in the DNLS model: this property is ensured
in the MMC setup by forbidding the change of arc, which would correspond to a shift of one or even two
sites of the breather itself.%
\footnote{We have implemented two additional rules which allow for breather diffusion, but
the coarsening exponent does not change (see later Fig.~\protect\ref{fig_coarsening}).}
 
In order to identify the breathers, it is necessary to introduce an absolute threshold $\theta$ to 
distinguish them from the background. Depending whether $a_i>\theta$ ($a_i<\theta$) a site is classified 
as a breather (background) and its amplitude denoted with $b_i$ ($g_i$). { A typical} example of the 
MMC evolution is shown in Fig.~\ref{fig_dynamics} (see the next section for a more accurate discussion,
where we show also that the threshold value $\theta$ is irrelevant in so far as it is large enough).

Let us now introduce the average and the variance of the two phases as,
\bea
b &=& \langle b_i \rangle  \qquad \qquad  \sigma_b^2 = \langle (b_i - b)^2 \rangle  \\
g &=& \langle g_i \rangle  \qquad  \qquad \sigma_g^2 = \langle (g_i - g)^2 \rangle \nonumber
\eea
where $\langle\dots\rangle$ is the spatial average, taken over all sites of the same family.

Imposing the conservation of $A$ and $H$, we obtain the {\it exact} relations,
\bea
A &=& (L-N)g +N b \\
H &=& (L-N) (g^2 + \sigma_g^2) + N(b^2 + \sigma_b^2)
\eea
where $N$ denotes the number of breathers. By defining 
the average amplitude and energy per site,
\be
a = {A\over L} , ~~~~~~
h = {H\over L} ,
\ee
we obtain 
\bea
a &=& \left( 1 - \rho \right) g + {b \rho} \label{exact_a}\\
h &=& \left( 1 - \rho \right) (g^2 + \sigma_g^2) + 
\rho (b^2 + \sigma_b^2)
\eea
where $\rho=N/L$ is the breather density (in the following, the average distance
$\lambda=\rho^{-1}$ will also be used).
In the limit $\rho \ll 1$,
\bea
g &=& a - b \rho \label{eq_linear} \\
h &=& a^2 - 2a b\rho + b^2\rho +\sigma_g^2 +\sigma_b^2\rho
\label{eq_quadratic}
\eea
where we have also used the information (obtained from numerics, see next section) 
that $b\gg g$, because $b$ increases in time while $g$ saturates.

Since the entire dynamics is invariant under a change of a scale, it is
convenient to rescale the amplitude to its average value $a$.
This is perfectly equivalent to assuming that $a=1$ (as we do from now on).
As a result,
\be
\rho = \frac{h -1 -\sigma_g^2}{ -2 b + b^2 + \sigma_b^2} \,.
\label{cons1}
\ee

In \cite{Rumpf2008}, on the basis of purely entropic arguments, it has been
found that for $h<2$ all breathers eventually disappear, otherwise one survives
(for $h>2$), accompanied by an infinite-temperature background,\comment{i.e.} characterized by a
Poissonian distribution { of the amplitudes}. Let us see, how such predictions manifest themselves in
the current setup.

From Eq.~(\ref{eq_linear}), for small densities, $g=1$ and 
$b \simeq ( h -1 -\sigma_g^2)\sqrt{\lambda}$ (recall that now $a=1$). Therefore,
the background has a finite amplitude, while the breather amplitude scales 
as the square root of breather average distance.
It is worth noting that $g \le 1$ and asymptotically $g\to 1$.
Therefore, for $\rho\ll 1$, since an infinite-temperature background corresponds to 
$\sigma_g^2 = g^2$, the request of a positive $\rho$ in Eq.~(\ref{cons1}) corresponds to
$h > 2$, the condition already derived in \cite{Rumpf2008}.

\section{Phenomenology}
In this section we illustrate the evolution, showing that it corresponds to a coarsening process.
We have worked with two classes of initial conditions:
(i) a fraction $f$ of equal-height breathers sitting on a homogeneous background characterized by
a uniform amplitude distribution within an interval of width $\delta a$ (ICa); (ii) a homogeneous
distribution of amplitudes characterized by the superposition of two exponential functions (ICb). 
We have verified that they give equivalent results, for the same value of $h$ (that is systematically
chosen to be larger than the critical value 2).

\begin{figure}
\centerline{\includegraphics[width=8 cm,clip]{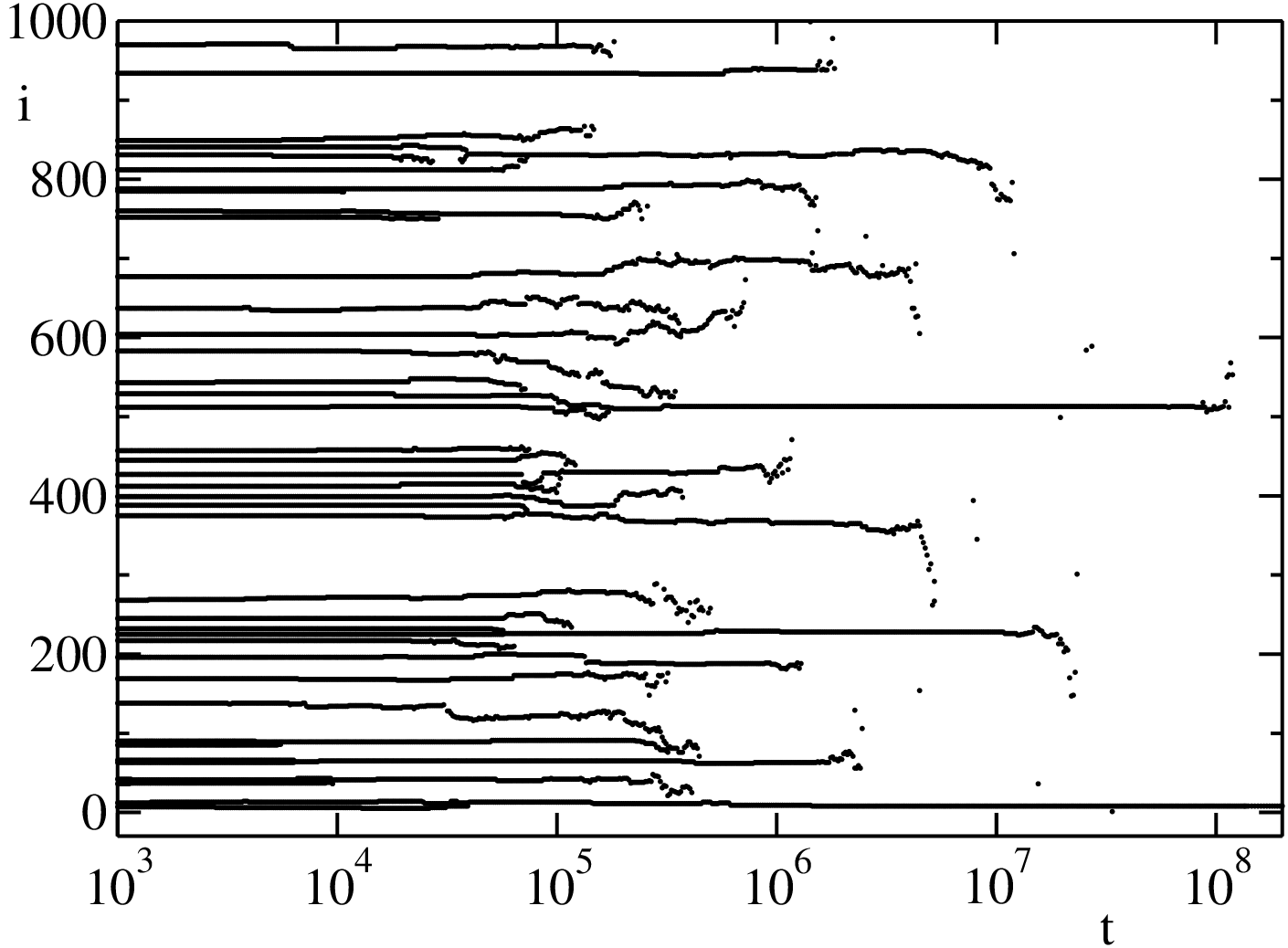}}
\caption{\comment{Breather dynamics for a chain of $1000$ sites with  $h=7.45$ and $\theta=12.5$.
The initial condition is of type
ICa with parameters $f=0.05$ and $\delta a=0.95$.}
{ Evolution of the local amplitude in a chain with $L=1000$ and $h=7.45$.
Breathers are identified by dots and correspond to the sites $i$ where $a_i>\theta$, with
$\theta=12.5$.
The initial condition is of type
ICa with parameters $f=0.05$ and $\delta a=0.95$.
}}
\label{fig_dynamics}
\end{figure}

In Fig.~\ref{fig_dynamics} we plot breathers positions as a function of time $t$, where
the time is measured in numbers of Monte Carlo moves divided by the system size ($L$).
The figure clearly shows the basic, qualitative features of dynamics: 
(i)~breathers do not typically move, but may diffuse just before disappearing under the threshold $\theta$
(i.e. when their amplitude becomes sufficiently small) ; 
(ii)~since the distinction breathers/background creates an artificial discontinuity,
a breather may disappear and then reappear for a short time;
(iii)~the density of breathers decreases in time (coarsening process), because breathers
gradually disappear. This process occurs when a breather goes below the threshold $\theta$. 

\begin{figure}
\centerline{ \includegraphics[width=8 cm,clip]{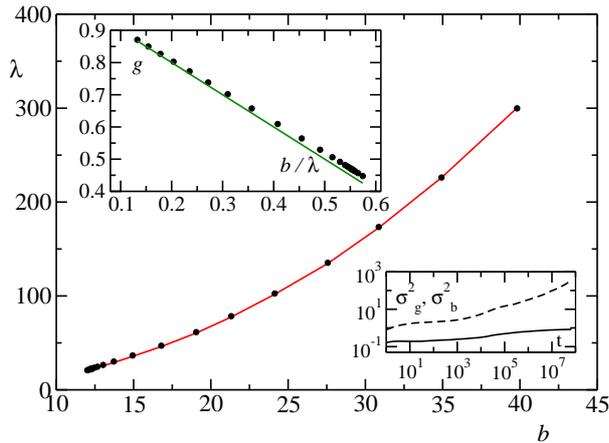}}
 \caption{Kinematics of the simplified DNLS model. 
 Black dots refer to direct simulations of a chain of $L=6400$ lattice sites, $h=7.23$
 and $\theta=8.45$
 (initial condition of type ICa with parameters $f=0.05$ and $\delta a=0.86$). 
 The solid red line is obtained according to Eq. (\ref{cons1}).
 {\it Upper inset:}  comparison of Eq. (\ref{eq_linear}) (solid green line) with simulations (black dots). 
 Bottom-right points 
 deviate from the expected trend as a consequence of a contribution of terms
 $\mathcal{O}(\lambda^{-1})$ in the early stages of coarsening.
 {\it Lower inset:} temporal evolution of the variances $\sigma^2_g,\sigma^2_b$.
 Asymptotically $\sigma^2_g$ (black solid line) converges
 to a finite value corresponding to the condition of infinite temperature background, 
 while $\sigma^2_b$ (black dashed line)
 increases as a consequence of the coarsening process.}
\label{fig_conservation_laws}
\end{figure}

The first quantitative analysis of the coarsening process is done by comparing direct simulations with
Eqs.~(\ref{eq_linear}), see Fig.~\ref{fig_conservation_laws}. The constraints derived from 
the conservation of total energy and amplitude  are found to be in good agreement with numerics
even in the region of small $\lambda$. Fig ~\ref{fig_conservation_laws} clearly shows that the 
coarsening process (i.e. the growth of $\lambda$) is directly related to an increase of the average
amplitude of the breathers, while the amplitude of the background remains finite. The same behavior
occurs also for the variances $\sigma^2_b$ and $\sigma_g^2$.

\begin{figure}[h]
\centerline{\includegraphics[width=8 cm,clip]{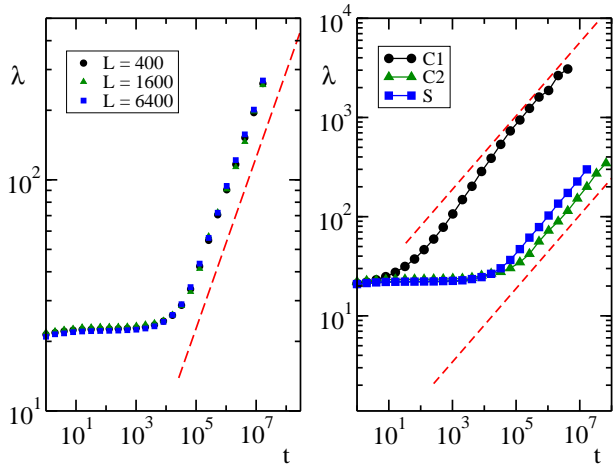}}
 \caption{{\it Left panel:} Average distance $\lambda$ versus time $t$ for three different
 system sizes $L=400,1600,6400$. 
 {\it Right panel:}
 Comparison of three different microscopic dynamics for a chain 
 of $N=6400$ lattice sites. 
Simulation parameters are the same of 
 Fig.~\ref{fig_conservation_laws}.}
\label{fig_coarsening}
\end{figure}

The next important quantitative aspect of dynamics concerns the coarsening law,
i.e. the time dependence of the distance between breathers, $\lambda(t)$, see
Fig.~\ref{fig_coarsening}. After an initial transient, $\lambda$ is found to grow in
time as $\lambda(t)\sim t^n$ with a coarsening exponent  $n=1/3$ that is independent of 
the system size (in fact the whole curve is independent of $L$, if
$L$ is large). 
Moreover, such asymptotic behavior is largely independent of the details
adopted to select a point in { the available phase space} \comment{$\mathcal{R}$}.
We have indeed considered two variations 
of the standard MMC algorithm (S). 
The first variant (C1) consists in selecting 
randomly a point either on the full circle or in the {\it union} of the three disconnected arcs. 
The second variant (C2) consists in partitioning the circle solutions in three symmetric arcs 
and selecting a point of the same arc as the initial configuration 
(also when the full circle would be available).  Altogether, the three algorithms
(S, C1, C2) can be classified according to their symmetry with respect to cyclic
permutations of the triplet (partial, full, absent, respectively).
It is quite remarkable to notice that the scaling behavior remains unchanged even when the 
breathers are allowed to diffuse in real space (setup C1). This is because the coarsening
does {\it not} proceed via coalescence (or annihilation) of the breathers: such a process is 
inconsistent with the simultaneous conservation of energy and mass. 

Dynamics, however, allows the exchange of matter between 
neighboring (and next neighboring) sites. In particular, breathers exchange matter with the 
underlying `sea', which acts as a mediator, so that breathers can effectively transfer matter between them.
As a consequence of that,
{ the breather amplitude fluctuates in time and it may go below threshold, and eventually be
absorbed by the background.}
\comment{breathers amplitude fluctuates and it may go
below threshold and the breather be eventually absorbed by the background.}
The key point of this evolution is the presence of a fast convergence towards local
equilibrium, where the background temperature becomes infinite as shown in the
next section. If two or more breathers are present, they interact via the background
(section~\ref{sec_two_breathers}) which allows them to exchange matter and acquire a finite 
life time which is the ultimate reason for the coarsening process, that will
be studied in detail in section~\ref{sec_SEP}.

Now let us analyze more precisely the exchange of matter between a (large) breather and the
surrounding background. This study will allow to understand that the `quantum' of 
transferred matter decreases with increasing breather amplitude.
In order to understand the dynamics in the presence of a large breather, let us consider a
triplet of consecutive sites and denote with $a$, $ya$ and $za$, the corresponding amplitudes 
in decreasing order from the largest to the smallest one, so that $1>y>z$.
If\,\footnote{This expression generalizes the condition $y,z<{1\over 4}$ for $y=z$, 
mentioned below Eqs.~(\protect\ref{dyn_rule}).}  
$1+y^2+z^2> 2(y+z+yz)$ the possible solution belongs to 3 distinct arcs. 
This is indeed the case when a breather is contained in the triplet since $ya$ and $za$ are both
much smaller than $a$, i.e. $y,z\ll 1$. Under this assumption ($y,z\ll 1$), it can be
easily shown that the rotation angle $\theta$ belongs to the interval $[-\sqrt{3}y,+\sqrt{3}z]$ 
and the new amplitude values are
\bea
a' &=&  \left[ 1 + \frac{\theta}{\sqrt{3}}(z-y)-\frac{\theta^2}{3} \right] a \nonumber\\
y' &=&  y+ \frac{\theta}{\sqrt{3}}(1-z)+\frac{\theta^2}{6}
\label{en_con} \\
z' &=&  z- \frac{\theta}{\sqrt{3}}(1-y)+\frac{\theta^2}{6}\nonumber
\eea
The largest variations are the opposite contributions $\pm\theta/\sqrt{3}$ which affect 
$y$ and $z$, respectively. As a result, the process corresponds, to leading order, to a
transfer of matter between two low-amplitude sites.  This means that
the breather is, in a first approximation, transparent to a propagation of matter.
In order to quantify the interaction of the breather with the background, we
have to consider the second order terms. Although the mass exchange is negligible
(of order $1/a$), the energy exchange is finite, no matter how big is $a$.

\begin{figure}[h]
 \centerline{\includegraphics[width=8 cm,clip]{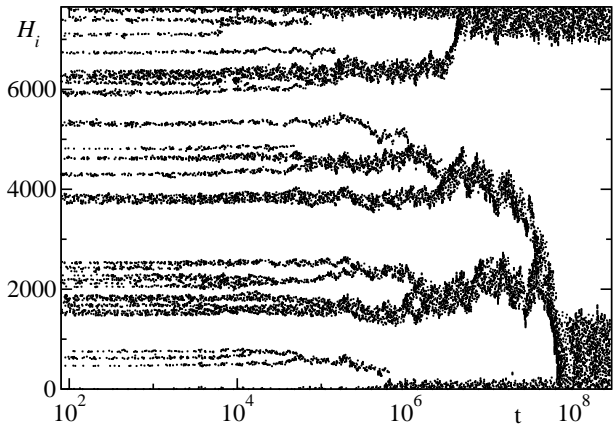}}
 \caption{Energy diffusion for a chain of 1000 sites and $h=6.75$. The initial condition is of type
 ICb with exponential rates $1.725$ and  $0.138$.}
\label{fig_dynamics2}
\end{figure}

Therefore, { we can conclude that it is neither} the position nor the mass of the
breather which performs a random motion, but rather its energy. This is preliminarily
confirmed by introducing a representation, where the total energy $H_i$ from site $1$ to site $i$ is plotted
for all $i$ versus time. The wide white regions visible in Fig.~\ref{fig_dynamics2} record the presence of 
breathers, while the dark areas correspond to sequences of background sites. The dark areas
coalesce, until the entire ``space" is split into a single black region (the homogeneous background) and a
single white region (one breather). The random fluctuations of the white areas signal the exchange of
energy between neighboring breathers.

\section{Fast relaxation}
\label{sec_one_breather}
In this section we discuss the relaxation to equilibrium of a single breather sitting on top of
a generic background. We analyze the process starting from an initial condition that is as
far as possible from the asymptotic state, i.e. from the highest possible breather amplitude $b$, 
compatible with the given energy and mass densities. This condition 
{(IC1 for later reference)} is achieved by selecting 
a constant background (i.e. zero-temperature) with amplitude $g$ such that $b^2+(L-1)g^2=hL$ 
and with $(L-1)g+b=L$. In the large $L$ limit, $g=1$ and $b = \sqrt{(h-1)L}$.

\begin{figure}[h]
\centerline{ \includegraphics[clip,width=8.cm]{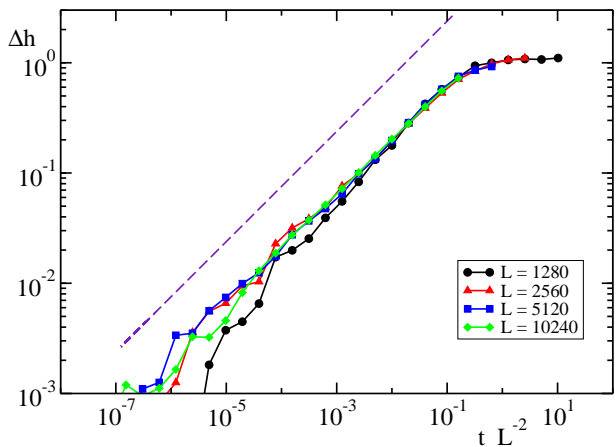}}
 \caption{ Scaling properties of the energy loss $\Delta h(t)$ of a single breather relaxing on
initially flat background for different system sizes $L$. The purple dashed line indicates a slope 1/2. 
Simulations are performed selecting initial conditions of type IC1 with $b=2\sqrt{L}$ (that corresponds
to a total energy density $h=5$).
}
 \label{LD_br}
\end{figure}

As soon as the system is let free to evolve, an energy-transfer
$\Delta H(t) = \langle b^2(0)-b^2(t) \rangle$ and a mass-transfer
$\Delta A(t) =\langle b(0) - b(t)\rangle$ set in, which contribute to 
decrease the breather amplitude.
Since $b$ is and remains on the order of $\sqrt{L}$, the energy $\Delta H(t)$ is an
extensive quantity (and one should more properly look at the intensive observable 
$\Delta h(t)= \Delta H(t)/L$), while $\Delta A(t)$ is subextensive, so that the
density of mass in the background is substantially unchanged. In practice, the transferred
energy contributes to increase background fluctuations.
Accordingly, one can equivalently monitor either $\Delta h(t)$ or the fluctuations 
$\sigma_g^2$. 
In Fig.~\ref{LD_br} we plot the evolution
of $\Delta h(t)$ for various system sizes and fixed $h$. A nice data collapse is
observed after introducing the scaling function
\begin{equation}
 \Delta H = L G(t/L^2), 
\label{eq:scalforce}
\end{equation} 
where $G(u)\approx \sqrt{u}$ for $u\ll 1$ and $G(u)\approx 1$ for $u\gg 1$.
In practice, up until times on the order of $L^2$, the energy transfer follows
a diffusive law $\Delta H = \sqrt{t}$ with a diffusion coefficient that is independent of $L$.
It is reasonable to conjecture that the energy absorption by the background is limited by
the diffusion over the background itself (this question will be discussed again
in a later section to justify the overall scaling behavior of the coarsening process).
The diffusion stops when the background reaches the maximally entropic state, i.e.
infinite temperature.

\begin{figure}[h]
\centerline{ \includegraphics[clip,width=8 cm]{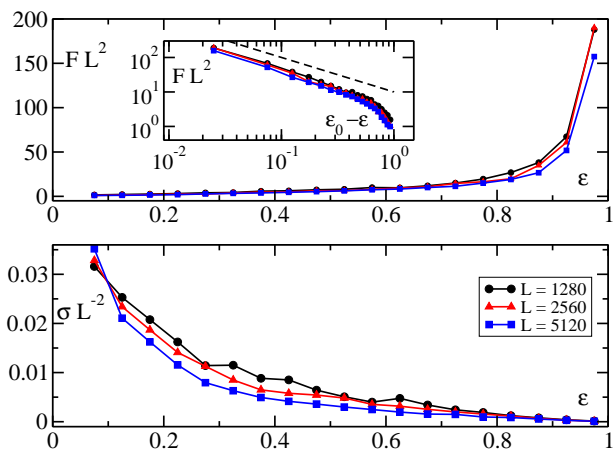}}
 \caption{{\it Upper panel:} Effective force $F(\eps)$ calculated via first passage times for three
  different sizes $L$.
 The zero of the horizontal axis corresponds to the equilibrium energy.
 The same quantity $F(\eps)$ is shown in the inset in terms of $(\eps_0-\eps)$, where $\eps_0=1$
 is the initial energy. The dashed line refers to a power law $1/(\eps-\eps_0)$.
 {\it Lower panel:} Standard deviation of the first passage times $\sigma$.
 $F$ and $\sigma$ have been calculated on a sample of $100$ independent initial conditions with 
 the same parameters of Fig.~\ref{LD_br} and $\delta \eps=0.05$.
}
\label{fig:efforce}
\end{figure}

It is convenient to approach the problem also in a different way by determining the first passage times
for the rescaled breather energy $\eps=b^2/L$.
In practice, we fix a series of equispaced thresholds $\eps_j$
($\delta \eps = \eps_{j+1}-\eps_j$) 
and determine the first time $t_j$ the energy crosses the $j$th threshold. Such times are then averaged 
over different realizations (see the angular brackets) to determine the effective force,
\begin{equation}
 F_j = \frac{\delta \eps}{\langle t_{j+1}-t_j\rangle} .
\end{equation} 
The results are plotted in Fig.~\ref{fig:efforce} (upper panel), where the origin of the $x$ axis is now chosen
at the  equilibrium value and the force is scaled by a factor $L^2$, 
consistently with Eq.~(\ref{eq:scalforce}). 
As a further check of consistency with the previous observation, notice that the force field
behaves, for $\eps$ approaching the state of maximal energy $\eps_0$ (equal to $1$ for the simulations
reported in Figs.~\ref{fig:efforce}), as
\begin{equation}
 F(\eps)L^2 \approx \frac{1}{\eps_0-\eps}
\end{equation} 
which both tells us that in $\eps_0$ there exists a barrier that cannot be overcome, and 
is consistent with the evolution of the breather energy.
In fact, in the continuum limit a solution of 
the deterministic dynamical equation
$\dot \eps = F(\eps)$ is
\begin{equation}
(\eps_0-\eps)  \approx \frac{\sqrt{t}}{L} 
\end{equation} 
(see the dashed line in Fig.~\ref{LD_br}). Therefore, we see that the pseudo-diffusive
law can be interpreted as the result of a divergence of the effective force.

We conclude this analysis by commenting about the amplitude of the stochastic fluctuations
that are plotted in the lower panel of Fig.~\ref{fig:efforce}. The diffusive force $\sigma$
is measured as the standard deviation of the first passage times. We see that $\sigma$
vanishes for $\eps=\eps_0$ where fluctuations can, indeed, only decrease the breather amplitude
and progressively increases until the fixed point is reached and we also see that
their amplitude scales as $L^2$. Altogether, this ``fast" relaxation process lasts a time of
order $L^2$.

Notice that one could have equivalently described the scenario by monitoring the average 
fluctuations $\sigma_g^2$ of the background. In such a case, we would have observed that the
variance of the background amplitude increases as in a diffusion process, converging
to a final state in which the  interface is made of random and independent heights.

\section{Slow relaxation and coarsening}
\label{sec_two_breathers}
%
%

\begin{figure}[h]
\centerline{\includegraphics[clip,width=8 cm]{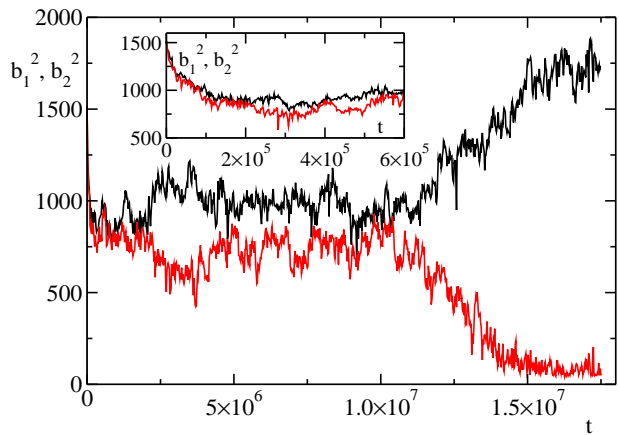}}
 \caption{
 Example of two breathers (with energies $b_1^2(t)$ and $b_2^2(t)$) relaxing in a chain with $L=1280$.
 The initial condition is characterized by a flat background and two
 equidistant breathers with amplitude $b_1(0)=b_2(0)=40$.
 The first part of the evolution (see the inset) corresponds to a
 an energy transfer
 from both the breathers to the background. Once the background has reached the infinite temperature
 point (for $t\gtrsim 5\times10^5$), the breathers start to exchange energy among themselves
 until one of them is completely blown  out.}
 \label{fig:2br}
\end{figure}

In the previous section we have studied how a single breather relaxes towards the equilibrium
state, where it has the optimal amplitude. 
In the presence of two (or more) breathers, the relaxation proceeds into two steps,
see Fig.~\ref{fig:2br}.
First,
the background converges towards the infinite-temperature state on a time scale that is analogous 
to that one studied in the previous section with reference to a single breather. Then,
a slow random exchange of energy between the breathers starts,  mediated by the background.
The energy of each breather performs a random walk with the constraint that the total breather energy 
is approximately constant. The process proceeds until the energy of a breather becomes so low that 
it is adsorbed by the background, and its energy has been transferred to the other breather(s). 
In this section we provide a characterization of such a slow dynamics by studying a simple setup that 
involves only two breathers. Finally, a straightforward generalization allows to explain the coarsening
exponent $n=1/3$.

In order to cut away the transient dynamics corresponding to the relaxation of the background,
the initial condition (IC2 for later reference) is now fixed by:
(i) generating an infinite-temperature 
background, where the amplitudes $a_i$ are i.i.d. variables, distributed according to the 
Poisson distribution $P(a_i) \propto \exp(-a_i/g)$~\cite{Rumpf2008}; (ii) adding two equidistant breathers 
with amplitude $b_1,b_2\gg g$ in a lattice of size $L$ with periodic boundary conditions.
If $L \gg 1$, the total mass is almost entirely contained in the background. Accordingly, 
mass conservation implies that the energy contained in the background is nearly conserved 
(provided that the background is at infinite temperature) and the energy contained in the two
breathers conserved as well. 
 
\begin{figure}[h]
\centerline{ \includegraphics[clip,width=8 cm]{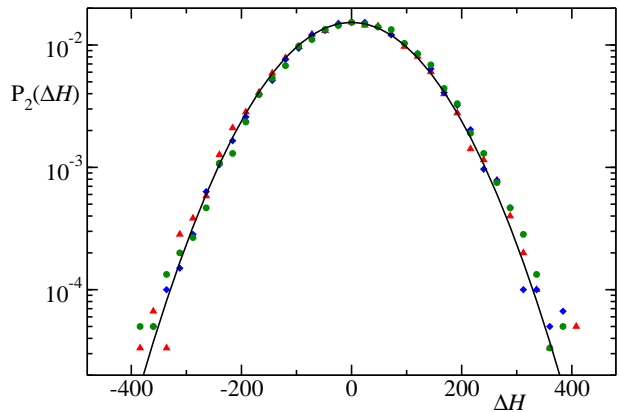}}
 \caption{Probability distribution of the energy transfer $\Delta H$ 
 of the first breather after a time
 $t=2^{19}$ in  a two-breather setup with initial conditions of type IC2.
 Three different configurations of initial breather amplitudes have been considered:
 $[b_1(0)=b_2(0)=b]$ (red triangles),
 $[b_1(0)=b_2(0)=b\sqrt{2}]$ (blue diamonds) and $[b_1(0)=b\sqrt{2}, b_2(0)=b]$ (green circles),
 with $b=100$.
 The black solid line refers to a Gaussian fit. Each data set is obtained from  a sample
 of $10000$ independent realizations of the MMC dynamics on a chain with $L=640$.
 }
\label{fig:distr}
\end{figure}

The first important property to point out is that the process of energy exchange between
the two breathers is independent of their amplitude. More precisely, in Fig.~\ref{fig:distr} we report the
probability distribution $P_2(\Delta H)$ of the energy transfer between two breathers after a time 
$t=2^{19}$, with $\Delta H=[b_1^2(0)-b_1^2(t)]$.
The distribution is very well fitted by a Gaussian function
with zero average and  appears unaffected by the choice of the initial
configuration $[b_1(0),b_2(0)]$. We can therefore conclude that such a process is determined only 
by the properties of the stationary background in between the two breathers.
From Fig. ~\ref{fig:distr} it is also clear that the only relevant information concerns the time evolution
of the variance $\sigma^2_2(t)$ of the distribution. 
According to stationarity, we can compute $\sigma^2_2(t)$ in terms of time averages, instead
of averages over independent initial conditions. In formulae, 
\begin{equation}\label{station}
 \sigma^2_2(t)\equiv\left\langle\left[b_1^2(0)-b_1^2(t)\right]^2 \right\rangle
 =\overline{\left[b_1^2(t+t_0)-b_1^2(t_0)\right]^2}\,,
\end{equation}
where the overbar denotes the average over time $t_0$  and $\langle\cdots\rangle$
denotes average over different initial conditions. Finally notice that stationarity holds 
as long as both breathers are present on the chain.
In the following of this section we will always explore this regime.

In order to perform a quantitative analysis of $\sigma^2_2(t)$, we have averaged the energy fluctuations
of a sample of $S$ independent trajectories
\footnote{
The average over independent trajectories is a practical numeric tool  for improving the statistics 
for large times $t$.} for different lattice sizes $L$.
Upon increasing $L$, we keep fixed the parameters $b_1$, $b_2$ and $g$ characterizing the initial 
condition IC2, so that 
the only change involves the distance between the two breathers.

\begin{figure}[h]
\centerline{ \includegraphics[clip,width=8 cm]{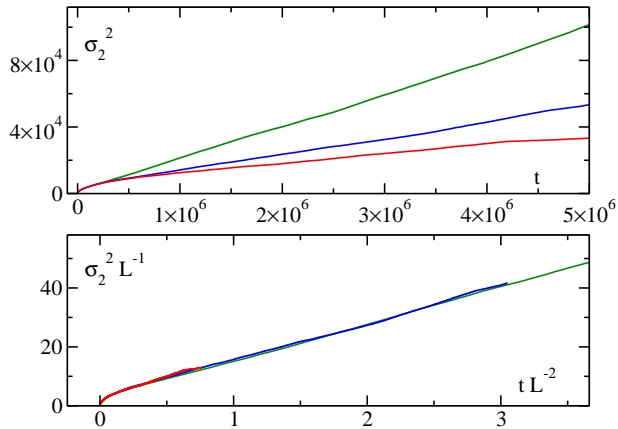}}
 \caption{
{\it Upper panel:} 
 Time evolution of the energy fluctuations $\sigma_2^2$ of two breathers sitting 
 on an infinite temperature background. Lines from top to bottom refer to lattice
 sizes $L=640$, $1280$, $2560$, respectively. Data are obtained from  a set 
 of $S=10$ samples 
 evolved for $3.4\times 10^7$ time units, starting from an initial condition of type IC2 with
 two breathers of amplitude $b=100$.
 Running averages
 are performed monitoring time differences from $dt=10^3$ to $dt=5 \times 10^6$ time units.
 {\it Lower panel:} Data collapse on the universal function $f_2(u)$ after the 
 rescaling $t\to t/L^2=u$ and $\sigma_2^2\to \sigma_2^2/L$. }
 \label{2br}
\end{figure}

The results reported in Fig. \ref{2br} show a growth in time of the energy fluctuations.
In particular,
for large enough times, $\sigma^2_2(t)$ has a linear profile, thus indicating
the existence of a diffusive law. The associated diffusion constant is however  inversely
proportional to the system size $L$ (and therefore also to the spatial separation of the breathers,
equal to $L/2$).
As a consequence, the larger the chain is, the slower the energy diffusion process is.
A data collapse is finally  obtained by rescaling energy and time respectively by 
$L$ and $L^2$.


\begin{figure}[h]
\centerline{ \includegraphics[clip,width=8 cm]{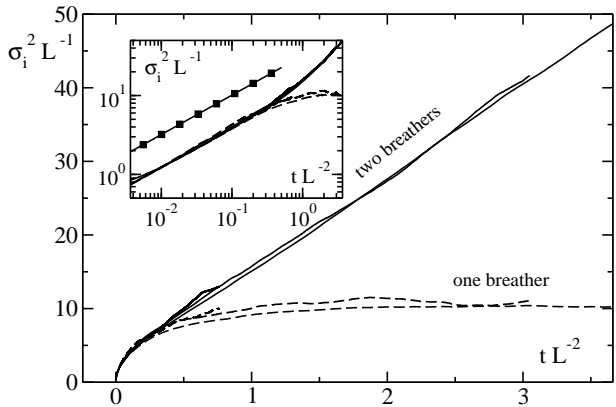}}
 \caption{Comparison of the universal functions $f_1(u)$ and $f_2(u)$ in presence 
 of one breather (dashed lines) and two breathers (solid lines), respectively. The functions  
 $f_i(u)$ are extracted  by means
 of the rescaling $t\to t/L^2=u$ and $\sigma_i^2\to \sigma_i^2/L$ for three different sizes
 $L=640, 1280, 2560$.
 Simulations are performed using the same parameters specified for Fig. \ref{2br}.
 The inset shows the behavior of $f_i(u)$ for small $u$. The square-solid line
 is a power law with exponent $1/2$.
 }
 \label{2+1br}
 
\end{figure}

A more general description of the overall scenario can be obtained by comparing the 
results of Fig. \ref{2br} with the same kind of simulations performed with only
one breather in the chain, i.e. in a equilibrium regime. Energy fluctuations,
denoted with $\sigma_1^2(t)$,
are now expected to be {\it finite} for large enough times, with the characteristic
scaling $\mathcal{O}(L)$ with the system size.
Fig.~\ref{2+1br} points out the equivalence of the dynamic behavior in presence
of one {\it and} two breathers for small times. In this regime, in fact, a breather
cannot know about the existence of some other breather in the chain and behaves 
as if it were at equilibrium (local equilibrium). Only after a characteristic 
time $t_c\sim L^2$ a ``bifurcation'' occurs, separating the linear growth
of fluctuations in presence of two breathers from the saturation required 
by (global) equilibrium in presence of only one breather.

The results plotted in Fig.s~\ref{2br} and~\ref{2+1br} can be summarized by the
following scaling relations for the variances $\sigma_1^2,\sigma_2^2$ of the energy of
a single breather when the system is made up of one and two breathers, respectively:
\be
\sigma_i^2 = L f_i\left( \frac{t}{L^2}\right)
\ee
where $f_1(u)\approx f_2(u) \approx \sqrt{u}$ for $u\ll 1$, while at large $u\gg 1$,
$f_1(u) \approx 1$ and $f_2(u) \approx u$.
In particular, $\sigma_2^2 \approx t/L$ at large times, which gives a first
explanation of the coarsening
exponent $n=1/3$. In fact, during the coarsening process the energy of the breather scales
as $L$, where $L$ is now the distance between breathers. Therefore, the time to allow
a breather to disappear corresponds to $\sigma_2^2 \sim L^2$, i.e.
$t/L\sim L^2$, and finally $L \sim t^{1/3}$.

\section{Comparison with a Partial Exclusion Process (PEP)}
\label{sec_SEP}

In the previous sections we have provided a characterization of the simplified DNLS
stochastic model in the presence of localized excitations. The very existence of such localized states
has been related to the simultaneous conservation of energy and mass. If, for example, we suppressed 
one conservation law, we would recover a standard diffusion process, always relaxing to 
an homogeneous state, with no room for breathers.
On the other hand, the stochastic process generated by the MMC rule turns out to be
nontrivial even at a microscopic level, since the available phase space
depends nonlinearly on the local amplitudes. From a macroscopic point of view, this property
leads to a nonlinear Fokker-Planck equation.
The question is therefore whether it is possible to further simplify the DNLS model, keeping all the 
essential features of its dynamics. This possibility might allow to obtain a more
rigorous explanation of the coarsening exponent $n=1/3$.

\begin{figure}[h]
\centerline{ \includegraphics[clip,width=8 cm]{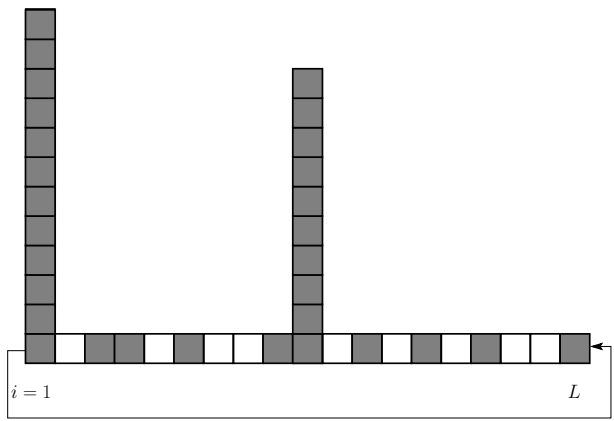}}
 \caption{The simplified model: two columns of bricks (grey squares) interact exchanging
 particles through a channel modeled by a partial exclusion process (PEP) with periodic boundary
 conditions. Each site of the channel
 can sustain  no more than one particle at a time. Particles evolve according to a standard diffusion 
 algorithm endowed with the exclusion rule.}
 \label{lego}
\end{figure}

{ In this section we show that a simple partial exclusive process (PEP) can reproduce the
coarsening observed in the DNLS context. The continuous variable $a_i$ is replaced 
by a discrete, integer amplitude $h_i$ which represents the number of particles on site $i$.
The evolution rule is purely stochastic: once an ordered pair of neighbouring sites $(i,j)$ 
($j=i\pm 1$) has been randomly selected, the transformation $(h_i,h_j)\to (h_i-1,h_j+1)$ 
is made if and only if $h_i > 0$ and $h_j \ne 1$. In practice, one can divide the lattice
into two parts: (i) breather-sites ($h_i>1$) which can freely exchange particles
with the surrounding environment; (ii) background ($h_i=0,1$) which behaves as in a standard
exclusion process. As soon as any given height $h_i$ decreases down to 1, the breather is 
irreversibly absorbed by the background.
Therefore, we can expect this model to exhibit a coarsening dynamics,
which will be compared to the dynamics of the simplified DNLS model.

In the previous sections it proved useful to study the relaxation of two-breather states
towards a final configuration characterized by a single breather. 
We are now going to do the same for the PEP model: see Fig.~\ref{lego}, where a configuration
with with breathers is sketched.
}

\comment{
In this section we compare the DNLS model  with a discrete-diffusion model based 
on a Partial Exclusion Process (PEP). Fig.~\ref{lego} shows a configuration in a system
of length $L$ and periodic boundary conditions, where two ``breathers'' interact via two channels.
The breathers are columns of discrete bricks of { height one}, so that
the total height of the column coincides with the number of bricks it is made of.
The columns can exchange particles by means of a SEP channel, where the occupation number 
can be either 0 (empty site) or 1 (occupied site).  The dynamics of the system is 
generated by a standard diffusion algorithm satisfying the exclusion rule. The sites
occupied by the columns represent an exception to the exclusion rule, as the 
occupation number can be greater than 1. If the height of a column becomes 1, then the column
becomes equivalent to the other sites of the channel. 
From that moment on, the standard exclusion rule is applied also to the original site of 
the column. Notice that, unlike the DNLS model which is characterized by two independent 
conservation laws, the dynamics of the PEP chain is characterized by one quantity alone, the 
total mass along the chain. On the other hand, the PEP model also has an additional constraint 
on the height of the background: this is the reason why it may be a useful, simplified model, 
for the original MMC dynamics.
}

\begin{figure}[h]
\centerline{ \includegraphics[clip,width=8 cm]{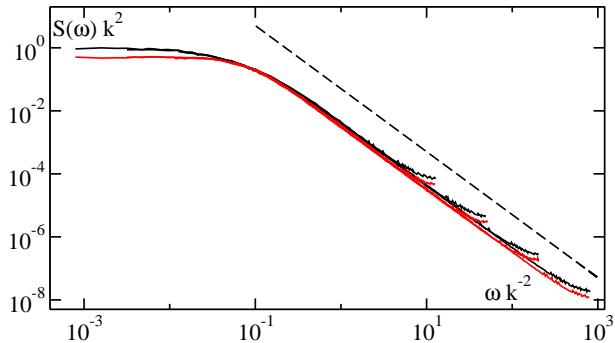}}
 \caption{
 Power spectra $S(\omega)k^2$ versus $\omega k^{-2}$ of the MMC dynamics (black solid line)
 and the PEP model (red solid line) for low-k Fourier modes, with $k=2\pi m/L$ and $m=\{1,2,4,8\}$.
 Data refer to MMC (PEP) chains with length $L=256$  in an infinite temperature state with average 
 occupation number $g=0.5$ $(\tilde\rho=0.5)$.
 The black dashed line is a power law with exponent $-2$. 
}
 \label{Spectr}
\end{figure}

We start by investigating the transport properties of the background field in
both models with the help of the power spectrum of the long-wavelength Fourier modes.
In order to study qualitatively similar dynamic regimes, we have chosen two initial
conditions with the same density of particles, in a condition of maximum entropy,
i.e. infinite temperature. In the MMC model, this requirement amounts to a Poissonian initial 
distribution of the amplitudes~\cite{Rumpf2008} with average $g$, while for the PEP model
the same regime corresponds to an average occupation number $\tilde\rho=0.5$. 
Accordingly, we have chosen $g=\tilde\rho=0.5$. The results are plotted in Fig.~\ref{Spectr}:
as expected the power spectrum of the PEP model is well fitted by a Lorentzian
distribution, a clear evidence of diffusion. A diffusive behavior in the
MMC model is not so straightforward, as its microscopic rule is much more complicated
than a standard diffusion algorithm, but this is what we find also in this case.
Moreover, the power spectra of the two models nicely overlap in a wide region of the Fourier
space and prove the validity of  the PEP model as an effective reduction of the MMC algorithm.
On the other hand, the slight gap in the low frequency region suggests a different behavior of
the (static) diffusion constants associated to the two dynamics.

\begin{figure}[h]
\centerline{ \includegraphics[clip,width=8 cm]{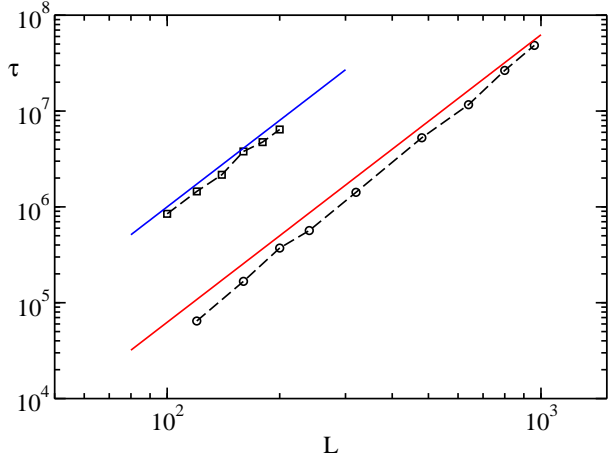}}
 \caption{
 Average lifetime $\tau$ of a couple of columns exchanging particles through a PEP channel 
 with length $L/2$. The columns have equal initial height $h_0$ linearly increasing with the system size
 $L$ as $h_0= kL$. Simulations have been performed choosing $k=1$ (open squares) and $k=1/4$ (open circles).
 The initial configuration of the channel corresponds to an infinite temperature state
 with average occupation number $\tilde\rho=0.5$. 
 Solid lines are defined by the equation $\tau=k^2 L^3$ with $k=1$ for the upper blue line and 
 $k=1/4$ for the lower red one.
}
 \label{ScalSEP}
\end{figure}

Having ascertained that perturbations propagate across the background diffusively in both 
models, we analyze the dynamics in the presence of breathers.
In the PEP model, we proceed by determining the mean lifetime $\tau$ of a couple of columns 
separated by two channels of length $L/2$ ($L$ being the total size of the system).
The quantity $\tau$ is defined as the average time that is necessary for one of the 
two columns to be destroyed (i.e. to reach a unitary height), supposing that the columns 
have initial heights $h_0=kL$, where $k$ is a positive constant. 
Fig.~\ref{ScalSEP} clearly shows that $\tau$ scales with the system
size as a power law, $\tau\sim L^3$. This scaling law is the same as for the original MMC
model and it is responsible for the characteristic coarsening process of the breathers 
with an exponent $n=1/3$.

Let us now evaluate the coarsening time for the PEP model and compare it with numerical results shown in
Fig.~\ref{ScalSEP}. If $h_0=kL$ is the initial breather height, the typical
time $\tau$ for the disappearance of one breather is given by the relation $h_0^2 = \tau/\Delta t$,
where $\Delta t$ is the typical time for exchanging one particle between the two breathers.
If the system were composed by the two neighboring breathers only, $\Delta t=1$, 
but in our case, once the ``emitting" breather has been chosen, the exchange becomes effective 
only if the move (breather)$\to$(neighbouring site) is allowed and if the diffused particle 
reaches the other breather before being re-absorbed by the emitting breather. The move 
(breather)$\to$(neighbouring site) occurs with probability $1/2$. which is the 
average occupancy of background sites. The diffusion to the other
breather occurs with probability $2/L$~\cite{Redner}, where $L/2$ is the distance between breathers. Therefore
$\Delta t = L$ and
\be
(kL)^2 = \frac{\tau}{L}
\ee
so that 
\be
\tau = k^2L^3
\ee

\section{Summary and conclusions}

In this paper we have thoroughly studied a microcanonical Monte Carlo model which approximately
reproduces the evolution of a DNLS equation for large mass densities
($a\gg1$). In the so-called negative temperature region (which corresponds to an energy density
$h>2a^2$), the spontaneous formation of breathers is observed
(see Figs.~\ref{fig_dynamics} and \ref{fig_dynamics2}). 

Our studies reveal that, in agreement with the entropic arguments put forward in
Ref.~\cite{Rumpf2007}, the system converges towards a stationary state composed of a single
breather in equilibrium with an infinite temperature background. The evolution is characterized
by two time scales.  Over ``fast" times (on the order $L^2$), the background equilibrates at infinite temperature,
while a population of breathers forms to store the ``excess'' energy that cannot be contained
in the background. This phenomenon can be attributed to the existence of two positive-defined conserved quantities 
(energy and mass) which scale with a different power of the local variable (quadratic and linear, 
respectively): as a result, breathers can contain extensive amounts of energy, while marginally contributing
to the mass which is, instead, essentially stored in the background. This regime has been first investigated
in the simplest possible setup: a single breather sitting on a
flat background. The data in Figs.~\ref{LD_br} and \ref{fig:efforce} clearly show the presence of a time
scale $L^2$.

On a longer time scale $L^3$, the evolution is characterized by a coarsening process, characterized by
an exchange of energy among the breathers, until they progressively disappear, leaving just one alive.
Coarsening occurs on a purely stochastic basis, because a breather may loose some of its
energy, which diffuses along the background possibly reaching a neighboring breather. 
Accordingly, the breather height fluctuates and, when a breather goes below threshold,
it is ``adsorbed" by the infinite temperature background and it cannot appear again.
The coarsening exponent $n={1\over 3}$ (see Fig.~\ref{fig_coarsening}) is a consequence of three
facts. First, conservation of mass and energy implies that the breather energy 
(not breather height, i.e. mass) performs a random walk. 
Second, conservation of mass and energy implies that the energy of a breather
scales as the distance $\lambda$ between breathers.
Third and finally, the elementary time scale for exchanging mass between 
breathers is set by the inverse of the probability that a ``quantum" of mass
released by a breather is able to attain the neighboring breather
instead of going back. This probability scales as $1/\lambda$.

{ In order to give a more firm basis to the derivation of $n={1\over 3}$, the furtherly
simplified PEP model has been introduced and analysed. Such a model shows remarkable similarities
with the original DNLS dynamics: (i) the condensation onto a single site appears above a critical value  
(defined by the mass density $\tilde \rho=0.5$ in the PEP model and by the energy density $h=2$ in
the simplified DNLS); the same scaling exponent is found in the two models. 
It is, however, necessary to recall a crucial difference: the presence of two rather than just
a single conservation law in the simplified DNLS model. The very same difference exists with other 
models where a similar scenario can be found: the Kinetic Ising Model with conserved 
magnetization~\cite{Kinetic} and a class of zero-range processes~\cite{zrp_godreche}.}

\comment{
The PEP model, which has been used to give more firm basis to the
derivation of $n={1\over 3}$, has some resemblance to the Kinetic Ising
Model with conserved magnetization~\cite{Kinetic} (which has the same coarsening
exponent). It is, however, worth stressing that such parallel between
PEP and simplified DNLS is limited to the coarsening process and it is
the conservation of mass and energy which allows its ``validation".
In other words and in simple terms, we might say that the simplified
DNLS model has two conservation laws and we use them to 
pass to the furtherly simplified PEP model. This remark is just to avoid
wrong shortcuts between DNLS and Kinetic Ising Model.} 

{ Finally, going back to the original DNLS model, it is worth recalling
two major simplifications introduced in this paper:
(i) the interaction energy between neighbouring sites has been neglected 
(the MMC model indeed corresponds to the high mass-density limit of the 
DNLS equation); (ii) dynamical effects are disregarded (the model is purely
stochastic). The first limitation could be removed by
reintroducing the interaction term in the original Hamiltonian.
As a result, one would be, however, forced to account also
for the phase dynamics (absent in the MMC model) with the
related difficulty (impossibility) of
deriving explicit microcanonical transformations. 
This is, nevertheless, a route that would be worth to explore,
especially to check the robustness of the coarsening process
herein investigated.
 
As for possible dynamical effects, one cannot exclude that
the dynamical arrest of the coarsening observed in \cite{iubini2013} 
is partly due to the weak coupling of the rapidly rotating
breathers with the background which makes energy and
mass exchanges even slower than in our stochastic process.}

%
%


\bibliographystyle{spphys}       
\bibliography{coarsNOdoi}   


\end{document}